\documentclass[
 reprint,
 superscriptaddress,
 amsmath,amssymb,
 aps,
]{revtex4-2}

\usepackage{float}
\makeatletter
\let\newfloat\newfloat@ltx
\makeatother

\usepackage{algorithm}
\usepackage{algpseudocode}
\usepackage{graphicx}% Include figure files
\usepackage{bm}% bold math
\usepackage{xcolor}
\usepackage{siunitx}
\begin{document}

% \preprint{APS/123-QED}

\title{Hierarchical organization of bursty trains in event sequences}

\author{Takayuki Hiraoka}
\affiliation{Department of Computer Science, Aalto University, Espoo 00076, Finland}%
\author{Hang-Hyun Jo}%
\email{h2jo@catholic.ac.kr}
\affiliation{Department of Physics, The Catholic University of Korea, Bucheon 14662, Republic of Korea}%

\date{\today}% It is always \today, today,

\begin{abstract}
Temporal sequences of discrete events that describe natural and social processes are often driven by non-Poisson dynamics. In addition to a heavy-tailed interevent time distribution, which primarily captures the deviation from a Poisson process, a heavy tail in the distribution of bursty train sizes is frequently observed, which implies the presence of higher-order temporal correlations that extend beyond interevent times. Here, we study empirical event sequences from different domains to show that the bursty trains in these processes are hierarchically structured across different timescales, and that such hierarchical organization gives rise to the higher-order temporal correlations. We propose a dynamic algorithm that generates event sequences with hierarchical structures with arbitrary precision. The algorithm successfully reproduces the features of real-world phenomena, implying the presence of memory mechanisms embedded in system dynamics across multiple timescales. 
\end{abstract}

\maketitle

%\tableofcontents

\section{Introduction}

One of the intriguing properties we encounter in human behavior as well as in biological and geological phenomena is temporal heterogeneity, commonly referred to as bursty dynamics~\cite{Karsai2018Bursty, Jo2023Bursty}. When a process is represented by a sequence of timestamped discrete events, burstiness is primarily characterized by a heavy tail in the interevent time (IET) distribution~\cite{Barabasi2005Origin, Bak2002Unified, Corral2004Longterm, Barabasi2005Origin, deArcangelis2006Universality, Vazquez2006Modeling, Bedard2006Does, Bogachev2007Effect, Malmgren2008Poissonian, Malmgren2009Universality, Kemuriyama2010Powerlaw, Wu2010Evidence, Tsubo2012Powerlaw, Jo2012Circadian, Kivela2015Estimating, Gandica2017Stationarity, Karsai2018Bursty}, as opposed to the exponential IET distribution for Poisson processes. In addition, higher-order temporal correlations, which are not captured by IET distributions, have been studied in terms of correlations between two successive IETs~\cite{Goh2008Burstiness, Baek2008Testing, Jo2018Limits}. More recently, Karsai et al.~\cite{Karsai2012Universal} introduced the notion of bursty trains for capturing correlations between an arbitrary number of consecutive IETs, which has been successfully applied to a wide range of empirical examples~\cite{Karsai2018Bursty, Yasseri2012Dynamics, Jiang2013Calling, Kikas2013Bursty, Wang2015Temporal, Jo2020Bursttree}. Bursty trains, or \emph{bursts}, are defined as clusters of successive events that follow one another within a given timescale $\delta$. The number of events in the burst is called burst size and denoted by $b$. In many empirical processes, the burst size distribution exhibits a power-law tail as
\begin{equation}
    P_\delta(b) \propto b^{-\beta_\delta}
    \label{eq:Pb}
\end{equation}
for a wide range of timescales $\delta$, where $\beta_\delta$ is the power-law exponent~\cite{Karsai2018Bursty}. If the IET sequence of such a process is randomly shuffled to destroy correlations between IETs, then it can be seen as a renewal process~\cite{Feller2009Introduction}; $P_\delta(b)$ for a renewal process decays exponentially for any IET distribution~\cite{Karsai2012Universal}. Therefore, the heavy tail in the burst size distribution indicates higher-order temporal correlations extending over multiple events.

To account for such correlations observed in empirical processes, several phenomenological models have been proposed. Karsai et al.~\cite{Karsai2012Universal} introduced a two-state model that incorporates memory reinforcement mechanisms. In this model, the system stochastically alternates between a normal state and an excited state, while generating events at different rates in each state. Reinforcement is applied to the times between events within the same state as well as to the state transitions. Later, Jo et al.~\cite{Jo2015Correlated} studied a variant of self-exciting point processes~\cite{Adamopoulos1976Cluster, Utsu1995Centenary, Ogata2006Space} that does not require a clear-cut distinction between two states. Instead, memory effects are taken into account by defining a time-varying event rate as the sum of excitations induced by past events that decay in time as a power law. Lee et al.~\cite{Lee2018Hierarchical} proposed a generative model with explicit hierarchical levels to show a power-law IET distribution but a stretched exponential distribution of burst sizes. However, none of these model provides full control over the distributions of IETs and burst sizes; thus they are insufficient to fully reproduce the empirical distributions observed in real-world data.

In contrast to the above generative approaches, Jo~\cite{Jo2017Modeling} investigated the correlation structure by first generating a sequence of uncorrelated IETs for a given IET distribution and then carefully permuting IETs to implement the higher-order temporal correlation such as heavy-tailed burst size distributions at several timescales. Although this static approach is effective in producing a sequence with a desired property, it is not sufficient to reveal the mechanisms underlying the dynamics.

More recently, Jo et al.~\cite{Jo2020Bursttree} introduced the burst-tree decomposition method, which maps an event sequence onto a rooted tree that represents the hierarchical structure of bursts. A burst tree describes how individual events are included in increasingly larger bursts, ultimately resulting in a burst containing all events in the process, as the timescale $\delta$ continuously increases from $0$ to the maximum IET in the sequence. The structure of a burst tree can be summarized by a burst-merging kernel dictating which bursts are merged as the timescale increases. The model that generates event sequences using the given burst-merging kernel has been systematically studied~\cite{Jo2020Bursttree, Birhanu2025Bursttree, Birhanu2025Maximum}. Yet, the interpretation of the burst-merging kernel in relation to the generative mechanism remains unclear, requiring further research.

In this paper, we propose an alternative approach to exploring the generative mechanisms behind higher-order temporal correlations in event sequences, focusing on the hierarchical burst structure over multiple timescales. We begin with an empirical analysis of natural and social phenomena to demonstrate that these processes are characterized by hierarchical self-similar structures. Next, we present a generative algorithm that can accurately reproduce any hierarchical structure; by means of numerical simulations, we show that this hierarchical structure is responsible for the heavy-tailed burst size distribution that spans a wide range of timescales. Finally, we discuss how higher-order temporal correlations are embedded in the dynamics. 

\section{Methods}

Let us consider an event sequence specified as an ordered set of timings of $n$ discrete events, denoted by $\mathcal{E}=\{t_i\}_{i=1,\ldots,n}$. By defining the interevent time (IET) as $\tau_i \equiv t_{i+1}-t_{i}$, we obtain from $\mathcal{E}$ the IET sequence $\mathcal{T}=\{\tau_i\}_{i=1,\ldots,n-1}$. For a given timescale $\delta$, a burst is defined as a set of events such that the IETs between any two consecutive events in the burst are less than or equal to $\delta$, while the first and last events of the burst are separated from events outside the burst by IETs greater than $\delta$~\cite{Karsai2012Universal}. The bursts at timescale $\delta$ constitute a partition of $\mathcal{E}$ and are ordered in time for any $\delta$. The number of events in a burst is called the burst size, and the size of the $j$th burst at timescale $\delta$ is denoted by $b_{\delta,j}$. By enumerating the size of each burst in time order, we have the burst size sequence $\mathcal{B}_{\delta}=\{b_{\delta,j}\}_{j=1,\ldots,m_\delta}$, where $m_\delta$ denotes the number of bursts at timescale $\delta$. Note that the sum of burst sizes is the same as the number of events, namely, $\sum_j b_{\delta, j}=n$ for any $\delta$. In the following, we omit the subscripts and represent the burst size by $b$ whenever it is not ambiguous. 

Here, we are interested in how bursts detected at different timescales are related to each other. Let us consider two consecutive bursts at timescale $\delta'$ that are separated by an IET $\tau>\delta'$. These bursts will merge to become a larger burst if the timescale increases to $\delta\geq\tau$. The size of the burst at timescale $\delta$ is the sum of the sizes of the bursts at timescale $\delta'$. Precisely, each burst size $b_{\delta,j}$ in $\mathcal{B}_{\delta}$ is obtained as the sum of the sizes of consecutive bursts in $\mathcal{B}_{\delta'}$, and the number of such bursts at timescale $\delta'$ is denoted by $k_{j}$:
\begin{align}
    b_{\delta, j}=\sum_{j'=s_{j}+1}^{s_{j}+k_{j}} b_{\delta',j'}, 
    \label{eq:k_define}
\end{align}
where $s_{j}$ is the number of bursts at timescale $\delta'$ that precede the $j$th burst at timescale $\delta$, namely,
\begin{align}
    s_{j}=\sum_{u=1}^{j-1} k_{u}
\end{align}
with $s_1=0$. Here we call $k_{j}$ the merging number, since it is the number of bursts at the smaller timescale $\delta'$ that are merged to make up one burst at the larger timescale $\delta$. The merging number sequence $\mathcal{K}_{\delta \delta'} = \{k_{j}\}_{j=1,\dots,m_\delta}$ thus fully describes the relationship between the burst size sequences at timescales $\delta$ and $\delta'$. Similarly to the notation for burst sizes, we simply use $k$ to denote the merging number whenever unambiguous.
    
\begin{figure}[!t]
\includegraphics[width=\columnwidth]{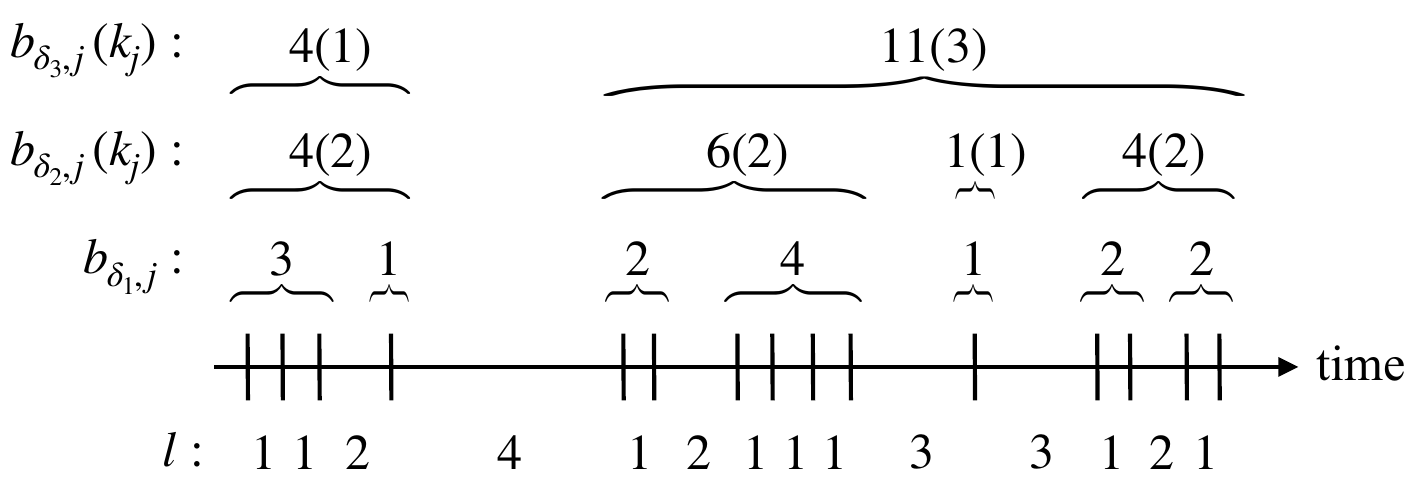}
\caption{Schematic diagram for the hierarchical structure of bursty trains at multiple timescales. Each vertical line in the time axis represents an event. $b_{\delta,j}$ denotes the size of the $j$th burst in the burst size sequence at the timescale $\delta$. Each $b_{\delta_l,j}$ is the sum of multiple consecutive $b_{\delta_{l-1},j'}$s, the number of which is called a merging number and denoted by $k_{j}$. We have assumed that $\delta_0<\tau_{\rm min}\le \delta_1$, with $\tau_{\rm min}$ denoting the minimum interevent time (IET), leading to $b_{\delta_0,j}=1$, hence $b_{\delta_1,j}=k_{j}$ from Eq.~\eqref{eq:k_define}. The number $l$ below each IET in the time axis denotes the level determining the range of the corresponding IET.}
\label{fig:scheme}
\end{figure}

As the timescale increases, bursts at smaller timescales are hierarchically merged into bursts at larger timescales, as illustrated in Fig.~\ref{fig:scheme}. The dendrogram representing this hierarchical clustering is called \emph{burst tree} and its structure is shown to be able to be associated with any IET distribution~\cite{Jo2020Bursttree}. Our aim in this paper is to study the structure of burst trees by focusing on the merging number distributions for several pairs of timescales. In a previous work, the merging numbers were assumed to be homogeneous~\cite{Jo2017Modeling}; we reexamine this assumption by studying the merging number distributions in empirical processes.

In the following analysis, we consider timescales between the minimum $\delta$ for which the maximum burst size exceeds 100 (i.e., $\max\{b_{\delta,j}\} \geq 100$) and the maximum $\delta$ for which the number of bursts are more than 100 (i.e., $m_\delta \geq 100)$ to ensure sufficient statistics of burst sizes. We will refer to this range of timescales as relevant timescales. 

We estimate the power-law exponents of distributions of burst sizes and merging numbers using the maximum likelihood estimator, and compare the goodness of fit of the power-law distribution and the exponential distribution by the likelihood ratio test~\cite{Clauset2009Powerlaw, Alstott2014Powerlaw}.

\section{Empirical results}

To uncover the structure of empirical processes by means of the distributions of IETs (derived from $\mathcal{T}$), burst sizes ($\mathcal{B}_\delta$), and merging numbers ($\mathcal{K}_{\delta \delta'}$), we analyze three datasets from different domains. 

The first dataset consists of neuronal firing obtained from electrophysiological recordings of pyramidal cell activity in rat hippocampus~\cite{Grosmark2016Diversity, GrosmarkDiversity}. During recording, the test animals underwent two four-hour rest/sleep periods interrupted by a 45-minute session in which they explored a maze room that was novel to them. The spikes in activity of each neuron constitute an event sequence.

The second dataset derives from Wikipedia. Wikipedia is an online encyclopedia whose articles are collectively contributed by volunteer editors. The edit history of each article and each editor can be retrieved publicly. We analyze preprocessed data derived from the English Wikipedia data dump on October 2, 2015~\cite{Jo2020Bursttree, Choi2021Individualdriven, 2015English}. We use the timestamps of the edits made by frequent editors, regardless of the articles they edited, to create event sequences.

The third dataset comes from earthquake observation in Southern California, United States, which represents a comprehensive archive of seismic data in the region~\cite{CaliforniaInstituteofTechnologyandUnitedStatesGeologicalSurveyPasadena1926Southern}. To ensure data consistency, we only use seismic events between January 1, 1993 to December 31, 2020. Each event is recorded with the time of occurrence, magnitude, and location (longitude, latitude, and depth) of the epicenter. We disregard the magnitude and depth information, and cluster the two-dimensional distribution of events on the longitude-latitude plane using \mbox{DBSCAN~\cite{Gan2007Data}}. By choosing appropriate parameter values (we set $\epsilon = \ang{0.07} \simeq 7\,\mathrm{km}$ and minimum number of points to 35, although the results are robust for a wider range of parameters), two large clusters, one in the south and one in the north, are obtained. The south cluster contains 330 thousand seismic events and corresponds to earthquakes occurring on the southern segment of the San Andreas Fault, the San Jacinto Fault, the Elsinore Fault, and the Los Angeles Basin. The north cluster consists of 150 thousand seismic events occurring along the Garlock Fault, the Kern Canyon Fault, and in the Indian Wells Valley. We consider the series of earthquakes in each cluster as an event sequence. 

\begin{figure*}[!t]
\includegraphics[width=\linewidth]{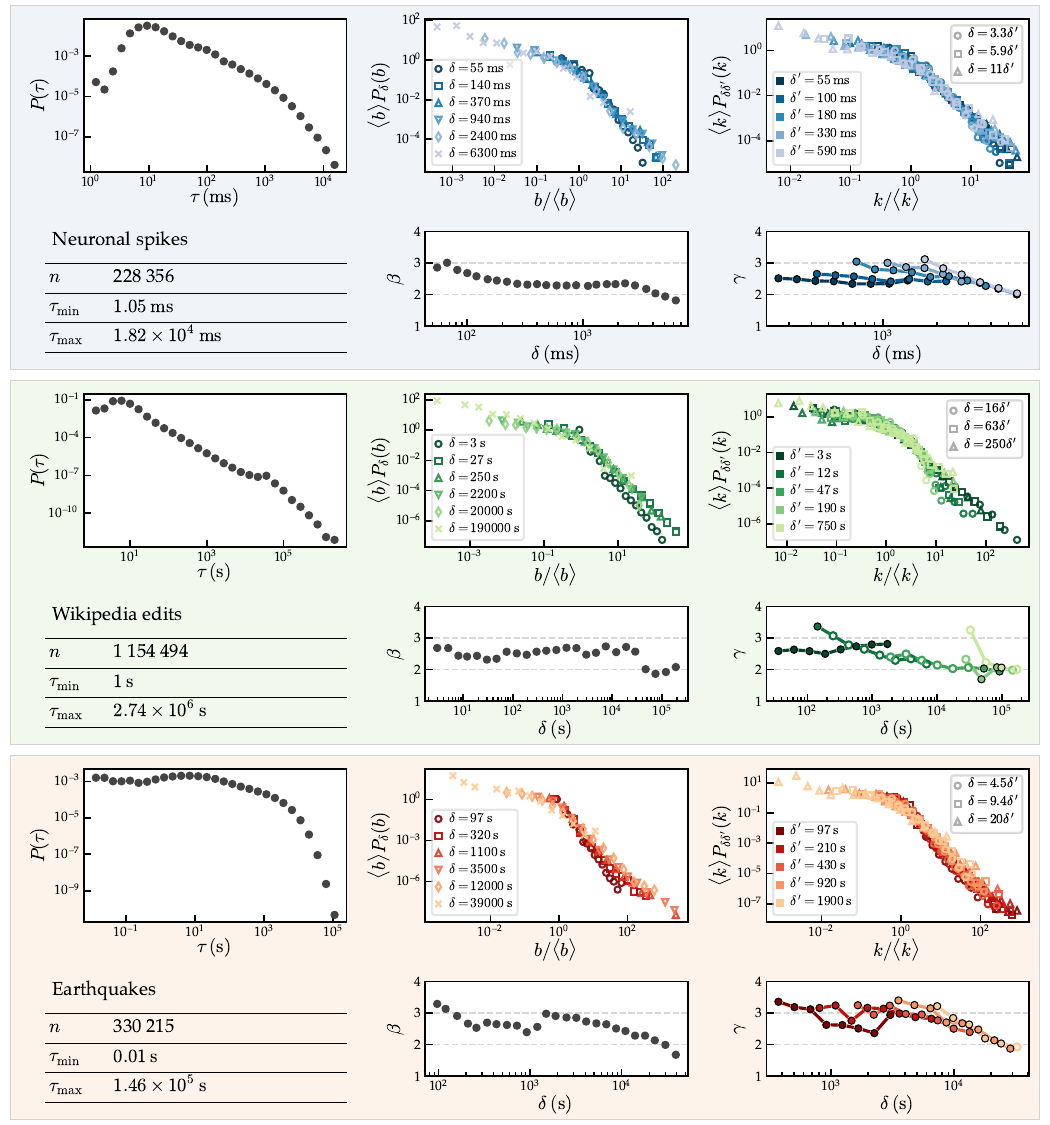}
\caption{Summary of event sequence analysis of empirical data: spikes of a rat neuron (top), edits made by a Wikipedia editor (middle), and earthquakes in Southern California (bottom). For each colored section, the top row shows the distributions of interevent times (left), burst sizes (center), and merging numbers (right). For the burst size and merging number distributions, the maximum likelihood estimators of the power law exponents are shown at the bottom. Fitted exponents are indicated by filled circles if the likelihood ratio test supports a power law distribution stronger than an exponential distribution with significance $p < 0.05$, and by open circles otherwise. The standard errors of the estimated exponents are smaller than the symbols and therefore not shown. The values of $\delta$ and $\delta'$ span the entire range of relevant timescales.}
\label{fig:empirical}
\end{figure*}

The results of the analysis are summarized in Fig.~\ref{fig:empirical}. Here, we show the results for a representative event sequence for each dataset: unit 9 in the \emph{Achilles-10252013} session in the neuronal spike recordings, the most frequent editor in the Wikipedia dataset, and the south cluster in the earthquake datasets. Other sequences in the datasets show similar patterns.

Several observations are in order. First, the range of IETs in each process extends across several orders of magnitude. Although not all the IET distributions are characterized by power laws, such strong heterogeneity of IETs is a hallmark of bursty processes. Second, we confirm previous findings that burst size distributions $P_\delta(b)$ are heavy-tailed across different timescales. In fact, our analysis shows that the estimated value of the power-law exponent fitted to the empirical distribution remains stable for several decades of timescales; namely, the exponent values are stable around $\beta_{\delta} \simeq 2.35$ for neuronal spikes, $\beta_{\delta} \simeq 2.50$ for Wikipedia edits, and $\beta_{\delta} \simeq 2.60$ for earthquakes. Third, and most interestingly, we find that the merging number distributions $P_{\delta \delta'} (k)$ also exhibit heavy tails for various timescale pairs. Moreover, the rescaled distributions for different pairs of $\delta$ and $\delta'$ collapse on top of each other, implying that the structure of the burst tree is self-similar across hierarchical levels or timescales. This is supported by fitting power-law tails; although the estimated exponents are not as stable as in burst size distribution, they remain relatively constant across a wide range of timescales. We express the merging number distribution with a power-law exponent $\gamma_{\delta\delta'}$ as follows:
\begin{align}
 P_{\delta \delta'}(k)\propto k^{-\gamma_{\delta\delta'}}.
 \label{eq:Pk}
\end{align}

The reason why the value of $\beta$ in Eq.~\eqref{eq:Pb} appears to be independent of the timescale $\delta$ can be understood by means of the result in Eq.~\eqref{eq:Pk}. Let us assume the following power-law tails for the burst size distribution at timescale $\delta$ and $\delta' < \delta$:
\begin{align}
    &P_{\delta}(b_\delta)\propto b_\delta^{-\beta_\delta},\\
    &P_{\delta'}(b_{\delta'})\propto b_{\delta'}^{-\beta_{\delta'}}.
\end{align}
Considering Eq.~\eqref{eq:k_define} and the power-law distribution of $k$ in Eq.~\eqref{eq:Pk}, one gets the scaling relation between $\beta_\delta$ and $\beta_{\delta'}$ as~\cite{Jo2013Contextual}
\begin{align}
    \beta_\delta =\min\{(\gamma_{\delta\delta'}-1)(\beta_{\delta'}-1)+1,\gamma_{\delta\delta'},\beta_{\delta'}\}.
    \label{eq:exponent_relation}
\end{align}
Note that the above scaling relation has been derived by assuming that summands in Eq.~\eqref{eq:k_define} are statistically independent of each other, whereas positive correlations between successive burst sizes were reported in empirical analyses~\cite{Jo2020Bursttree}. Despite of such fact, let us explore the possibility of applying Eq.~\eqref{eq:exponent_relation} to our empirical results. Since $\gamma_{\delta\delta'},\beta_{\delta'}>2$ in most cases as shown in Fig.~\ref{fig:empirical}, we have
\begin{align}
    \beta_\delta=\min\{\gamma_{\delta\delta'},\beta_{\delta'}\}.
\end{align}
This agrees well with observations: In the case of neuronal spikes, for example, the value of $\beta_\delta$ is quite stable for the range of $\delta<3000$ ms, for which we observe $\beta_{\delta'}<\gamma_{\delta\delta'}$. For $\delta>3000$ ms, the decreasing tendency of $\beta_\delta$ seems to be driven by the decreasing $\gamma_{\delta\delta'}$. This argument can be applied to cases with other datasets. Overall, we find a flat behavior of $\beta_\delta$ when $\beta_{\delta'}<\gamma_{\delta\delta'}$, while the decreasing behavior of $\beta_\delta$ might be driven by a decreasing $\gamma_{\delta\delta'}$. 

\section{Dynamic generative algorithms}

Based on the empirical findings in the previous section, we propose a model that dynamically generates events one by one in a sequence with a desired hierarchical structure of bursts. We assume that timescales indexed by $l$ are discrete states (levels) that the system is in. The system transitions between levels while it generates events with different time intervals according to the level it is in. Each state $l$ is equipped with an independent memory mechanism, embodied by the variable $\kappa_l$, which keeps track of the number of events the system generates in that state and controls whether it stays in the current level or switches to another one. Precisely, the memory variable $\kappa_l$ represents the remaining number of bursts that needs to be generated at timescale $\delta_l$ to make up a single burst at timescale $\delta_{l+1}$, where $\delta_l$ strictly increases with $l$. Our model has two inputs, i.e., the merging number distributions $P_{\delta_{l+1}, \delta_l}(k)$ for $l=0,1,\ldots$, and the IET distributions $P_l(\tau)$ for $l=0,1,\ldots$.

Specifically, the algorithm is formulated as follows. Initially, the system is assumed to be at level $l = 0$ and the first event is generated at $t_1 = 0$. Given merging number distributions between consecutive timescales, $P_{\delta_{l+1}, \delta_l}(k)$, the system performs one of the following operations at each step based on its current level $l$:
\begin{enumerate}
    \item If $\kappa_l = 0$ or the value of $\kappa_l$ is not set, a number sampled from the merging number distribution $P_{\delta_{l+1}, \delta_l}(k)$ is set to $\kappa_l$.
    \item If $\kappa_l = 1$, $\kappa_l$ is decremented to zero and the system moves up to level $l + 1$.
    \item Otherwise, the $i$th event is generated at time $t_i=t_{i-1}+\tau$, where $\tau$ is drawn from the IET distribution for level $l$, $P_l (\tau)$. $\kappa_l$ is decremented by one and the system moves down to the lowest level $l = 0$.
\end{enumerate}
We repeat this procedure until the desired number of events $n$ is reached ($i = n$) or the desired duration $T$ is exceeded ($t_i \geq T$). In this manner, each burst at level $l + 1$ is composed of $k$ bursts at level $l$ for $l \geq 1$, where $k$ is drawn from $P_{\delta_{l+1}, \delta_l}(k)$, while the bursts at level $0$ are of size $1$. The algorithm is summarized as a pseudocode in Algorithm~\ref{alg:decremental}.

\begin{algorithm}[t]
	\caption{Decremental dynamic algorithm.} 
	\begin{algorithmic}
		\State $l \gets 0$
        \Comment{Current level}
		\State $i \gets 1$
        \Comment{Event index}
        \State $t_i \gets 0$
        \Comment {Time of the first event}
		\While{$i < n$}
        \Comment{Loop until $n$ events are generated}
			\If {$\kappa_l = 0$ or $\kappa_l$ is not set}
				\State $k \sim P_{\delta_{l+1}, \delta_l}(k)$
                \Comment{Sample merging number}
				\State $\kappa_l \gets k$
                \Comment{Reset memory variable}
			\ElsIf {$\kappa_l = 1$}
				\State $\kappa_l \gets \kappa_l - 1$
                \Comment{Decrement memory variable to zero}
				\State $l \gets l + 1$
                \Comment{Increment current level}
			\Else
				\State $\tau \sim P_l (\tau)$
                \Comment{Sample interevent time}
                \State $i \gets i + 1$
				\State $t_i = t_{i-1} + \tau$
                \Comment{Set event time}
				\State $\kappa_l \gets \kappa_l - 1$
                \Comment{Decrement memory variable}
				\State $l \gets 0$
                \Comment{Reset level to the lowest}
			\EndIf
		\EndWhile
	\end{algorithmic} 
    \label{alg:decremental}
\end{algorithm}

While this model is fully dynamic, one may argue that it is unphysical because the system anticipates its future behavior by observing the number of bursts it will generate at each level $l$ that will be grouped into a single burst at level $l + 1$. A more realistic model would rely only on the memory of the system's past behavior to control how the next event is generated, rather than injecting the memory of its future. To this end, we formulate a statistically equivalent model that only uses past information by considering a memory process in the same spirit as the approach taken in Karsai et al.~\cite{Karsai2012Universal}. In this model, the system makes a stochastic decision whether to add another burst at the current level $l$ that will be included in a higher-level burst at $l+1$ or to move itself up to level $l+1$. The probability of the former occurring, denoted by $q_{\delta_{l+1}, \delta_l} (\kappa_l)$, depends on the number $\kappa_l$ of bursts that have already been generated within the same higher-level burst.

\begin{figure}[!h]
\includegraphics[width=\linewidth]{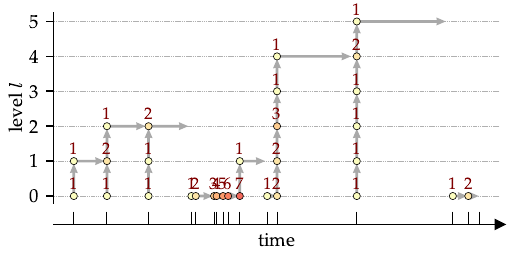}
\caption{Illustrative example of the incremental dynamic algorithm. Colored circles indicate the state of the system at each iteration, with the number on top of each circle denoting the value of $\kappa_l$. Rightward horizontal arrows represent the system sampling an IET and generating an event, after which it moves down to the lowest level. Upward vertical arrows represent transitions to the level above. The spike series at the bottom shows the generated event sequence.}
\label{fig:generative_schematic}
\end{figure}

The algorithm is initialized similarly as before: the system is at level $l = 0$, and the first event is generated at time $t_1 = 0$. In this algorithm, the memory variable is initialized as $\kappa_l = 1$ for each level $l$ the first time the system visits $l$. At each step, the system performs either of the following:
\begin{enumerate}
    \item With probability $q_{\delta_{l+1}, \delta_l} (\kappa_l)$, an event is generated at time $\tau$ after the last event in the sequence, where $\tau$ is drawn from the IET distribution for the current level $l$, $P_l (\tau)$. Memory variable $\kappa_l$ is incremented by one and the system moves down to the lowest level $l = 0$.
    \item With probability $1 - q_{\delta_{l+1}, \delta_l} (\kappa_l)$, memory variable $\kappa_l$ is reset to one and the system moves up to level $l+1$.
\end{enumerate}
Algorithm~\ref{alg:incremental} presents a pseudocode of this algorithm, and Fig.~\ref{fig:generative_schematic} shows a schematic example of the algorithm. The difference between this algorithm and the previous one is that the memory variable is incremented instead of decremented each time an event is generated; we will refer to the first algorithm as ``decremental'' and the second as ``incremental''.

\begin{algorithm}[t]
	\caption{Incremental dynamic algorithm.} 
	\begin{algorithmic}
		\State $l \gets 0$
        \Comment{Current level}
		\State $i \gets 1$
        \Comment{Event index}
        \State $t_i \gets 0$
        \Comment {Time of the first event}
		\While{$i < n$}
        \Comment{Loop until $n$ events are generated}
			\If {$\kappa_l$ is not set}
                \State $\kappa_l \gets 1$
                \Comment{Initialize memory variable}
            \EndIf
            \State $r \sim \mathcal{U}(0, 1)$
            \Comment{Draw a random number}
            \If{$r < q_{\delta_{l+1}, \delta_l}(\kappa_l)$}
                \State $\tau \sim P_l (\tau)$
                \Comment{Sample interevent time}
                \State $i \gets i + 1$
                \State $t_i = t_{i-1} + \tau$
                \Comment{Set event time}
                \State $\kappa_l \gets \kappa_l + 1$
                \Comment{Increment memory variable}
                \State $l \gets 0$
                \Comment{Reset level to the lowest}
            \Else
				\State $\kappa_l \gets 1$
                \Comment{Reset memory variable}
                \State $l \gets l + 1$
                \Comment{Increment current level}
			\EndIf
		\EndWhile
	\end{algorithmic} 
    \label{alg:incremental}
\end{algorithm}

The probability function $q_{\delta \delta'} (\kappa)$ controls how likely another burst at timescale $\delta'$ occurs within a higher-level burst at timescale $\delta$ given that $\kappa$ bursts have already occurred. It is referred to as the memory function and related to $P_{\delta \delta'}(k)$ as~\cite{Karsai2012Universal} 
\begin{equation}
    q_{\delta \delta'} (\kappa) = \frac{\sum_{k = \kappa + 1}^\infty P_{\delta \delta'}(k)}{\sum_{k = \kappa}^\infty P_{\delta \delta'}(k)}.
    \label{eq:q_kappa_define}
\end{equation}
The memory function encodes the entire information about the probability distribution, which is why the incremental and decremental algorithms can be made statistically equivalent. 

Any distribution bounded between $\delta_l$ and $\delta_{l+1}$ is suitable for the IET distribution $P_l (\tau)$ for level $l$; for example, we can use the uniform distribution 
\begin{equation}
    P_l(\tau)=\frac{1}{\delta_{l+1}-\delta_l},
\end{equation}
or the log-uniform distribution
\begin{equation}
    P_l(\tau)=\frac{1}{\tau (\log \delta_{l+1} - \log \delta_l)}.
\end{equation}
In the numerical calculation below, we use the uniform distribution.

The values of timescales $\delta_l \, (l=0, 1, 2, \dots)$ are arbitrary so long as they are nonnegative and increasing. Importantly, they do not affect the statistical property of the burst tree structure; it only controls the IET distribution of the generated event sequence. To demonstrate the proposed algorithm, we adopt three different series of timescales: 
linearly increasing timescales
\begin{equation}
    \delta_l = \delta_0 + wl, \, w > 0;
    \label{eq:lints}
\end{equation}
quadratically increasing timescales
\begin{equation}
    \delta_l = \delta_0 + wl^2, \, w > 0;
    \label{eq:quadts}
\end{equation}
and exponentially increasing timescales 
\begin{equation}
    \delta_l = \delta_0 w^l, \, w > 1.
    \label{eq:expts}
\end{equation}
We also assume for simplicity that the merging number distribution is independent of $l$, hence simply denote it by $P_{\delta_{l+1}, \delta_l}(k) = P(k)$, although it is easy to relax this assumption. In the following numerical demonstration, $P(k)$ is assumed to be a zeta distribution:
\begin{align}
    P(k)=\frac{1}{\zeta(\gamma) k^{\gamma}} 
    \label{eq:Pk_input}
\end{align}
for $k = 1, 2, \dots$, where $\zeta$ is the Riemann zeta function. The corresponding memory function is given by Eq.~\eqref{eq:q_kappa_define}:
\begin{equation}
    q_{\delta \delta'} (\kappa) = \left(\frac{\kappa}{\kappa + 1}\right)^{\gamma - 1}.
\end{equation}

\begin{figure*}[!t]
\includegraphics[width=\linewidth]{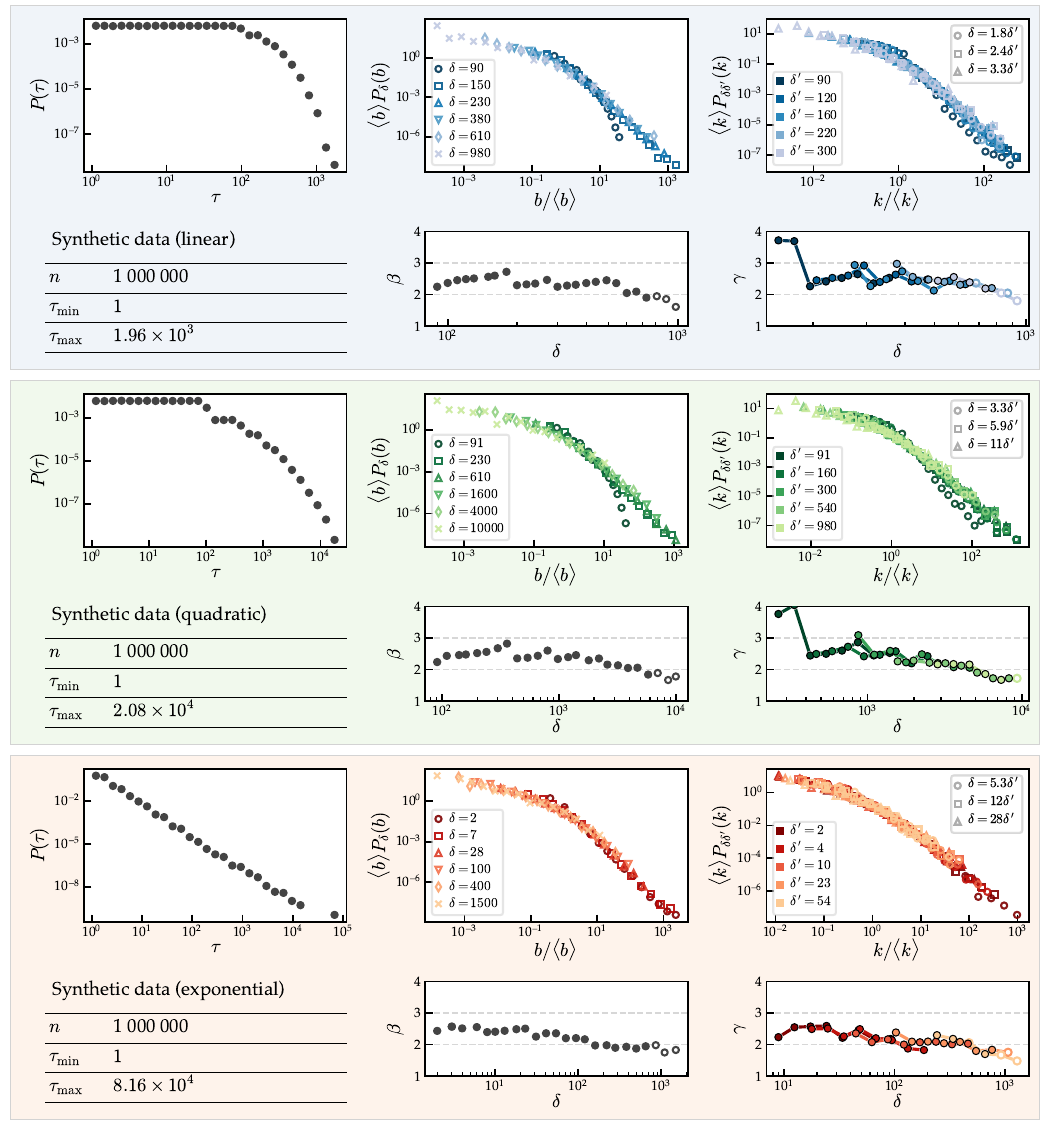}
\caption{Summary of event sequence analysis of synthetic data generated by the incremental dynamic algorithm with linear timescales (top), quadratic timescales (middle), and exponential timescales (bottom). Statistical analyses, panel arrangement, and symbol conventions are the same as in Fig.~\ref{fig:empirical}.}
\label{fig:synthetic}
\end{figure*}

Figure~\ref{fig:synthetic} summarizes the analysis of event sequences generated by the incremental dynamic algorithm. Here, we set the parameter values as $n = 10^6$, $\gamma = 2.5$, and $\delta_0 = 1$. We use three different series of timescales described above, with $w = 100$ for the linear [Eq.~\eqref{eq:lints}] and quadratic [Eq.~\eqref{eq:quadts}] series, and $w = 2$ for the exponential series [Eq.~\eqref{eq:expts}]. We omit the results for event sequences generated by the decremental algorithm since it produces statistically equivalent sequences as the incremental one.

For all of the three timescale series, the generated sequences show heavy-tailed distributions of burst sizes $b$ for a wide range of timescales $\delta$, indicating that the algorithm successfully reproduces higher-order correlations seen in real-world sequences. The distributions of merging numbers $k$ are also heavy-tailed across different pairs of timescales $\delta$ and $\delta'$, even though these timescales do not coincide with the preset series of timescales $\delta_l$. Notably, the power-law exponent values of the heavy tails in $P_{\delta}(b)$ and $P_{\delta \delta'}(k)$ are stable across timescales.

The IET distributions $P(\tau)$ show distinct shapes depending on the series of timescales $\delta_l$. Interestingly, some of these shapes are similar to those observed empirically. For instance, the IET distribution for the sequence generated using a linear timescale series has a similar shape to that for the earthquake sequence, while the power-law IET distribution for the sequence based on the exponential timescales is reminiscent of that for the data on Wikipedia edits. This implies the presence of multi-timescale memory mechanisms underlying real-world phenomena. That is, the shape of the IET distribution, together with the hierarchical burst organization, reveals multiple timescales inherent in the dynamics of the system, ranging from temporally coarse-grained (i.e., slow) to fine-grained (fast). 

\section{Conclusions}

In this paper, we have shown that real-world event sequences from various domains follow the same organizational principle: hierarchical burstiness in several decades of timescales. We have identified that this hierarchical structure is responsible for the universal observation of higher-order temporal correlations. The burstiness at different timescales implies a multiscale memory mechanism behind the dynamics of the system. We have introduced a dynamic algorithm with independent memory embedded in discrete timescales and demonstrated that it is able to reproduce the empirical observations. 

Our dynamic algorithm is general and flexible enough to accommodate any desired interevent time distribution and hierarchical organization of bursts in terms of merging number statistics. Therefore, we expect this model to serve as a canonical model for generating synthetic event sequences with high-order temporal correlations. Moreover, it provides insights into the workings of the underlying memory mechanisms, revealing how memory functions across different timescales in each process. In particular, the fact that the memory variable $\kappa_l$ is reset every time the system transitions to a higher level implies that memory at each timescale is not carried over between bursts separated by longer interevent times. Instead of maintaining extensive memory of past information, the system only remembers the number of successive bursts at each timescale until those bursts aggregate into a higher-level burst, while correlations that extend beyond that are retained in a coarser-grained way at longer timescales.

It is reasonable to speculate that the multiscale organization of an event sequence is associated with the spectrum of dynamical modes, ranging from fast to slow, which is suggested and/or evidenced in various systems in the literature~\cite{Bak2002Unified, Corral2004Longterm, Kiebel2008Hierarchy, Peng2010Integrated, Proekt2012Scale, Okuda2018Hierarchical, Palva2018Roles, Raut2020Hierarchical}. Identifying such mechanisms by fitting our model to empirical would be an interesting avenue of future research.

\begin{acknowledgments}
The authors thank Woo-Sik Son for providing us with the preprocessed dataset of the English Wikipedia. H.-H.J. acknowledges financial support by the National Research Foundation of Korea (NRF) grant funded by the Korea government (MSIT) (No. 2022R1A2C1007358).
\end{acknowledgments}


%\bibliography{h2jo-papers}% Produces the bibliography via BibTeX.

\end{document}